\documentclass[pra,aps,showpacs,superscriptaddress,twocolumn]{revtex4}
\usepackage{graphicx}
\usepackage{amsmath}
\usepackage{amsfonts}
\usepackage{amssymb}
\usepackage{epsfig}
\usepackage{color}

\begin{document}

\newcommand{\abs}[1]{\left\vert#1\right\vert}
\newcommand{\set}[1]{\left\{#1\right\}}
\newcommand{\bra}[1]{\left\langle#1\right\vert}
\newcommand{\ket}[1]{\left\vert#1\right\rangle}
\newcommand\braket[2]{\left.\left\langle#1\right|#2\right\rangle}
\def\I {{\rm 1} \hspace{-1.1mm} {\rm I} \hspace{0.5mm}}
\newcommand{\rosso}[1]{\color[rgb]{0.6,0,0} #1}

\title{Off-resonant entanglement generation in a lossy cavity}

\author{F. Francica}
\affiliation{Dip. Fisica, Universit\`a della Calabria, \& INFN -
Gruppo collegato di Cosenza, 87036 Arcavacata di Rende (CS) Italy}
\author{S. Maniscalco}
\affiliation{Department of Physics and Astronomy, University of Turku, FI-20014 Turku, Finland}
\author{J. Piilo}
\affiliation{Department of Physics and Astronomy, University of Turku, FI-20014 Turku, Finland}
\author{F. Plastina}
\affiliation{Dip. Fisica, Universit\`a della Calabria, \& INFN -
Gruppo collegato di Cosenza, 87036 Arcavacata di Rende (CS) Italy}
\author{K.-A. Suominen}
\affiliation{Department of Physics and Astronomy, University of Turku, FI-20014 Turku, Finland}

\date{\today}

\begin{abstract}
We provide an exact and complete characterization of the
entanglement dynamics of two qubits coupled to a common structured
reservoir at zero temperature. We derive the conditions to
maximize reservoir-induced entanglement for an initially
factorized state of the two-qubit system. In particular, when the
two qubits are placed inside a lossy cavity, we show that high
values of entanglement can be obtained, even in the bad cavity
limit, in the dispersive regime.  Finally we show that, under
certain conditions, the entanglement dynamics exhibits quantum
beats and we explain their physical origin in terms of the
interference between two different transitions coupling the
dressed states of the system.
\end{abstract}

\bigskip

\maketitle

\section{Introduction}
Quantum entanglement is the powerful resource lying at the root of
a new class of technologies based on the laws of quantum theory.
The coherent manipulation of quantum systems involves very
delicate procedures since the inevitable interaction with their
surroundings leads to a loss of information that causes both the
transformation of quantum superpositions into statistical mixtures,
a process called decoherence,  and the disappearance of quantum
entanglement in composite systems.

Recently, it has been shown that entanglement can be lost
completely in a finite time despite the fact that complete
decoherence only occurs asymptotically. This phenomenon, named
entanglement sudden death, has been theoretically predicted  by Yu
and Eberly \cite{yu}, and experimentally observed for
entangled photon pairs \cite{natu} and atomic ensembles
\cite{kimble}. Typically, entanglement sudden death occurs when
the two qubits interact with two independent environments as for
the case, e.g., of two entangled qubits placed inside two
different cavities. For such a configuration, a class of states
has been identified which do not experience a complete
entanglement loss despite the interaction with local vacuum
environments \cite{marek}. However,  for
finite temperature environments the sudden death occurs almost
independently of the initial state of the qubit pair \cite{al-q},
although with details that can depend on the amount of
non-Markovianity of the environments \cite{bellomo}. In this
context, a deeper understanding of the sudden death process has
been gained by looking at the quantum correlations shared by the
environments which show a sudden birth (though with a quite
counterintuitive timing) \cite{sainz,lopez}.

A completely different phenomenology emerges when the qubits
interact with the same environment. In this case, indeed,
entanglement can be created starting from a factorized state or it can
even revive after a sudden death. This is due to
the effective qubit-qubit interaction mediated by the common
reservoir
\cite{Benatti03,Benatti06,Oh06,Liu,An07,natali07,Dajka08,ficek}.
Many theoretical papers have studied reservoir-induced
entanglement in the Markovian regime, that is when the coupling
between the qubits and the environment is weak enough to neglect
the feedback of information from the reservoir into the system
(memoryless dynamics). An interesting extension to these
approaches, that  goes beyond the Born-Markov approximation, has
been presented in Ref. \cite{Hu}.

In a recent paper, we have studied the dynamics of two qubits
coupled to a common structured environment using an exact approach
that does not rely on the Born-Markov approximation \cite{Man08}.
We focused on the case in which the qubits were identical and
resonant with the cavity field, whose spectrum was modelled as a
Lorentzian. In this paper we extend our analytical approach to
describe the more general situation in which the qubit
frequencies are different and non-resonant with the main mode
supported by the cavity. Our new analytical results allow us to
characterize completely and exactly the entanglement dynamics for
a generic initial two-qubit state containing one excitation. We
study the time evolution of the entanglement and its dependence on
several parameters, all in principle adjustable in the
experiments: the relative coupling between the atoms and the
cavity field, the initial amount of entanglement, the frequency of
the qubits, the detuning from the cavity field, and the quality
factor of the cavity. In this way we determine the conditions to
achieve maximal reservoir-induced entanglement generation for an
initial factorized state of the qubits, and to minimize the loss of
entanglement for an initial entangled state.

Depending on the matching of the qubit frequencies, we will
distinguish two scenarios displaying different qualitative long
time behavior. If the two qubits have the same transition
frequency ($\omega_1=\omega_2$), a decoherence-free state
(subradiant state) exists \cite{palma,zanardi}. Due to the
presence of such a dark state, a non-zero asymptotic entanglement
can be obtained in this case. On the other hand, if the two qubits
have different transition frequencies ($\omega_1 \neq \omega_2$),
no subradiant state exists, so that the stationary entanglement
always vanishes. For the sake of brevity, we refer to these two
cases as subradiant and non-subradiant scenario, respectively.

One of our main results is the demonstration that high values of
reservoir-induced entanglement can be obtained  in the dispersive
regime even in the  bad cavity limit. In general, in this regime
the dynamics of the concurrence (that we employ to quantify
entanglement \cite{wootte}) is characterized by a quasi-regular
and quasi-periodical pattern since the cavity photon is only
virtually excited and therefore the two-qubit system is less affected
by the cavity losses. Finally, in the good cavity limit, we
predict the occurrence of quantum beats of entanglement and
explain their physical origin.

The paper is structured as follows. In Sec. II we present the
microscopic Hamiltonian model, for which the exact analytical
solution is presented in Sec. III, where we focus on the case in
which the spectrum of the environment is Lorentzian as, e.g., for
the electromagnetic field inside a lossy resonator. In Sec. IV
and V we present and discuss our main results by looking at the
entanglement dynamics in the subradiant and non-subradiant
scenarios, respectively, for different coupling regimes and
different initial states. Finally, Sec. VI contains summary and
conclusions.

\section{The model}

We study an open quantum system consisting of two  qubits coupled
to a common zero-temperature bosonic reservoir in the vacuum. The
Hamiltonian describing the total system is given by
\begin{equation}\label{eq:H}
    H=H_S+H_R+H_{\rm int},
\end{equation}
where $H_S$ is the Hamiltonian of the qubits system coupled, via
the interaction Hamiltonian $H_{\rm int}$, to the common
reservoir, whose Hamiltonian is $H_R$.

The Hamiltonian for the total system, in the dipole and
the rotating-wave approximations, can be written as (assuming
$\hbar = 1$)
\begin{eqnarray}
H_S &=& \omega_1 \sigma^{(1)}_+  \sigma^{(1)}_- + \omega_2
\sigma^{(2)}_+\sigma^{(2)}_-, \\
H_R &=& \sum_k \omega_k b_k^{\dag} b_k, \\
H_{\rm int} &=&  \left( \alpha_1 \sigma^{(1)}_+   +
\alpha_2\sigma^{(2)}_+ \right) \sum_k g_k b_k  + {\rm h.c.},
\label{eq:Hint}
\end{eqnarray}
where $b^\dag_k$, $b_k$ are the creation and annihilation
operators of quanta of the reservoir,  $ \sigma^{(j)}_{\pm}$ and
$\omega_{j}$ are the inversion operators and transition frequency
of the $j$-th qubit  ($j=1,2$); finally $\omega_k$ and $\alpha_j
g_k$ are the frequency of the mode $k$ of the reservoir and its
coupling strength with the $j$-th qubit.


Here, the $\alpha$'s are dimensionless real coupling constants
measuring the interaction strength of each single qubit with the
reservoir. In particular, we assume that these two constants can
be varied independently. In the case of two  atoms inside a
cavity, e.g., this can be achieved by changing the relative
position of the atoms in the cavity field standing wave. We denote
with $\alpha_T=(\alpha_1^2+\alpha_2^2)^{1/2}$ the collective
coupling constant and with $r_j=\alpha_j/\alpha_T$ the relative
interaction strength.

\subsection{Dynamics of the qubit system}

We assume that initially the qubit system and the reservoir are
disentangled. We restrict ourselves to the case in which only one
excitation is present in the system and the reservoir is in the
vacuum. In this case  the initial state for the whole system can
be written as
\begin{equation}
\ket{\Psi(0)} = \Bigl [ c_{01} \ket{1}_1\ket{0}_2 + c_{02}
\ket{0}_1\ket{1}_2\Bigr] \bigotimes_k
\ket{0_k}_R,\label{initialstate}
\end{equation}
where  $\ket{0}_j$ and $\ket{1}_j$ $(j=1,2)$ are the ground and
excited states of the $j$-th qubit, respectively, while
$\ket{0_k}_R$ is the state of the reservoir with zero excitations
in the mode $k$.

The time evolution of the total system, under the action of this
Hamiltonian,  is given by
\begin{eqnarray}
\vert \Psi (t) \rangle &=&  c_1 (t) \vert 1 \rangle_1 \vert 0
\rangle_2 \vert 0 \rangle_R + c_2(t)  \vert 0 \rangle_1 \vert 1
\rangle_2 \vert 0 \rangle_R+ \nonumber \\ &&+ \sum_k c_k (t) \vert
0 \rangle_1 \vert 0 \rangle_2 \vert 1_k \rangle_R, \label{eq:psi}
\end{eqnarray}
where $\vert 1_k \rangle_R$ is the state of the reservoir with
only one excitation in the $k$-th mode and $\vert 0 \rangle_R =
\bigotimes_k \ket{0_k}$.

The reduced density matrix describing the two-qubit systems,
obtained from the density operator $\vert \Psi(t) \rangle \langle
\Psi(t) \vert$ after tracing over the reservoir degrees of
freedom, takes the form

\begin{equation}
\label{eq:rhos} \rho(t) = \begin{pmatrix}
   0& 0 & 0 & 0 & \\ 0& |c_1(t)|^2   &  c_1(t)c_2^*(t)  & 0\\
 0& c_1^*(t)c_2(t)      & |c_2(t)|^2  & 0\\
   0&  0  & 0 & 1-|c_1|^2-|c_2|^2
\end{pmatrix}.
\end{equation}
The two-qubit dynamics is therefore completely characterized by
the amplitudes $c_{1,2}(t)$.

Introducing the $j$-qubit detuning from the mode $k$,
$\delta_k^{(j)} = \omega_j-\omega_k$, the equations for the
probability amplitudes take the form
\begin{eqnarray}
\dot{c}_j(t)&=&-i \alpha_j  \sum_k g_k e^{i \delta_k^{(j)}t}
c_k(t), \quad j=1,2\label{eq:cj} \\ \dot{c}_k(t)&=&-i  g^*_k
\left[ \alpha_1 e^{-i \delta_k^{(1)} t} c_1(t) +\alpha_2 e^{-i
\delta_k^{(2)}t} c_2(t) \right]. \label{eq:ck}
\end{eqnarray}

Formally integrating Eq.~(\ref{eq:ck}) and inserting its solution
into Eqs.~(\ref{eq:cj}), one obtains two integro-differential
equations for $c_{1,2}(t)$,

\begin{eqnarray}
\nonumber \dot{c}_1(t) &=& -\sum_k \int_0^t \! dt_1 \Big[ \,
\alpha_1^2
\abs{g_k}^2 e^{i\, \delta_k^{(1)}(t-t_1)}c_1(t_1) \\
&& +\, \alpha_1 \alpha_2 \abs{g_k}^2 e^{i\, \delta_k^{(1)}t}
e^{-i\, \delta_k^{(2)}t_1} c_2(t_1) \Big], \label{eq:Dotc1}
\end{eqnarray}
\begin{eqnarray}
\nonumber \dot{c}_2(t)&=&-\sum_k \int_0^t \! dt_1 \Big[\, \alpha_1
\alpha_2 \abs{g_k}^2 e^{i\, \delta_k^{(2)}t} e^{-i\,
\delta_k^{(1)}t_1}
c_1(t_1)\\
&& +\, \alpha_2^2 \abs{g_k}^2 e^{i\, \delta_k^{(2)}(t-t_1)}
c_2(t_1) \Big]. \label{eq:Dotc2}
\end{eqnarray}

In the continuum limit for the reservoir spectrum the sum over the
modes is replaced by the integral
$$
\sum_k \abs{g_k}^2 \rightarrow \int \! d\omega J(\omega),
$$
where $J(\omega)$ is the reservoir spectral density. In the
following we focus on the case in which the structured reservoir
is the electromagnetic field inside a lossy cavity. In this case,
the fundamental mode supported by the cavity displays a Lorentzian
broadening due to the non-perfect reflectivity of the cavity
mirrors. Hence the spectrum of the field inside the cavity can be
modelled as
\begin{eqnarray}
\label{eq:J} J(\omega) = \frac{W^2}{\pi} \frac{\lambda}{\left(
\omega - \omega_c \right)^2 + \lambda^2},
\end{eqnarray}
where the weight $W$ is proportional to the vacuum Rabi frequency
and $\lambda$ is the width of the distribution and therefore
describes the cavity losses (photon escape rate).

We now introduce the correlation function $f(t-t_1)$, defined as
the Fourier transform of the reservoir spectral density
$J(\omega)$,
$$
f(t-t_1)=\int \! d\omega J(\omega) e^{i(\omega_c-\omega)(t-t_1)},
$$
where $\omega_c$ is the fundamental frequency of the cavity.
In terms of the correlation function Eqs. (\ref{eq:Dotc1})-(\ref{eq:Dotc2}) become
\begin{eqnarray}
\dot{c}_1(t)&=&-\int_0^t \! dt_1 \left[ \, \alpha_1^2 \, c_1(t_1)
+
\alpha_1 \alpha_2 \, c_2(t_1) e^{-i \, \delta_{21} t_1} \right] \nonumber \\
 && \times f(t-t_1) e^{i\, \delta_1 (t-t_1)},
\label{eq:Dotc1f}
\end{eqnarray}
\begin{eqnarray}
\nonumber \dot{c}_2(t)&=&-\int_0^t \! dt_1 \left[ \alpha_1
\alpha_2 \, c_1(t_1) e^{i\, \delta_{21} t_1} + \alpha_2^2 \,
c_2(t_1) \right]\\
&&\times f(t-t_1) e^{i\, \delta_2 (t-t_1)}, \label{eq:Dotc2f}
\end{eqnarray}
where $\delta_j=\omega_j-\omega_c$ and $\delta_{21}=\omega_2-\omega_1 $.

Performing the Laplace transform of
Eqs. (\ref{eq:Dotc1f})-(\ref{eq:Dotc2f}) yields
\begin{eqnarray}
s \, \widetilde{c}_1(s)-c_1(0) &=& - \left[\, \alpha_1^2 \,
\widetilde{c}_1(s) + \alpha_1 \alpha_2 \, \widetilde{c}_2(s+i \,
\delta_{21})\, \right] \nonumber \\ && \times \widetilde{f}(s-i \,
\delta_1), \label{eq:LapTrc1f}
\end{eqnarray}
\begin{eqnarray}
 s \, \widetilde{c}_2(s)-c_2(0) &=&
- \left[\, \alpha_1 \alpha_2 \, \widetilde{c}_1(s-i \,
\delta_{21}) + \alpha_2^2 \, \widetilde{c}_2(s)\, \right]
\nonumber \\ && \times \widetilde{f}(s-i \, \delta_2).
\label{eq:LapTrc2f}
\end{eqnarray}
From the equations above one can derive the quantities
$\widetilde{c}_1(s)$ and $\widetilde{c}_2(s)$. Finally, inverting
the Laplace transform one obtains a formal solution for the
amplitudes $c_1(t)$ and $c_2(t)$. The main steps for deriving the
general solution are outlined in Appendix A. For specific forms of
the reservoir spectral density, as the one we consider in this
paper, it is possible to obtain simple analytic expressions for
these coefficients.

Before discussing the general features of the dynamics we notice
that, when the two qubits have the same transition frequency,
$\omega_1 = \omega_2$, a subradiant, decoherence-free state
exists, that does not decay in time. The existence of the
subradiant state does not depend on the form of the spectral
density and therefore on the resonance/off-resonance condition.
Such a state  takes the form
\begin{eqnarray}
\label{eq:psim} \ket{\psi_-} = r_2 \ket{1}_1 \ket{0}_2  - r_1
\ket{0}_1 \ket{1}_2.
\end{eqnarray}
When the two qubits have different frequencies, $\omega_1 \neq
\omega_2$, there is no decoherence-free state.

This simple consideration enables us to draw general conclusions
about the dynamics of entanglement for long times. Indeed, one can
observe two qualitatively different behaviors. In the subradiant
scenario, occurring for $\omega_1=\omega_2$, a subradiant state
exists and therefore that part of the initial entanglement stored
in $\ket{\psi_-}$ will be \lq trapped\rq \ for arbitrary long
times. In the non-subradiant scenario, when $\omega_1 \neq
\omega_2$, the subradiant state does not exist. Hence all 
initial entanglement will decay and is eventually lost for long
times.

We now derive the solution for the coefficients $c_1(t)$ and
$c_2(t)$ and study the entanglement dynamics discussing separately
the two cases outlined above.

\subsection{Subradiant Scenario}
For $\omega_1=\omega_2$ the analytical solution for the amplitudes  $c_1(t)$ and $c_2(t)$ takes a
simple form, with a structure analogous to the solution of the resonant case presented in Ref.  \cite{Man08},

\begin{eqnarray}
\!\!\! c_1(t) \!\!&=&\!\! \left[\, r_2^2 + r_1^2 \, {\cal E}(t)\,
\right]c_1(0) - r_1 r_2
\left[\, 1- {\cal E}(t)\, \right]c_2(0), \nonumber \\ \label{eq:c1Sc1} \\
 \!\!\! c_2(t)\!\!&=&\!\! - r_1 r_2 \left[\, 1- {\cal E}(t)\,
\right]c_1(0)+ \left[\, r_1^2 + r_2^2\, {\cal E}(t)\,
\right]c_2(0), \nonumber \\ \label{eq:c2Sc1}
\end{eqnarray}
with
\begin{eqnarray}
{\cal E}(t) = e^{- (\lambda -\, i \delta)\, t /2} \left[
\cosh \left ( \Omega t /2\right) + \frac{\lambda - i
\delta}{\Omega} \, \sinh \left( \Omega t /2 \right) \right], \nonumber \\
\label{eq:E}
\end{eqnarray}
where $\delta_1=\delta_2\equiv \delta$ and $\Omega= \sqrt{
\lambda^2 - \Omega_R^2-i 2 \delta \lambda}$, with $\Omega_R =
\sqrt{4 W^2 \alpha_T^2 + \delta^2}$ the \emph{generalized Rabi
frequency} and ${\cal R} = W \alpha_T$ the \emph{vacuum Rabi
frequency}.

As in the resonant case,
the state $\ket{\psi_-}$ does not evolve in time and the only
relevant time evolution is the one of its orthogonal superradiant
state
\begin{eqnarray}
\label{eq:psip} \vert \psi_+ \rangle = r_1 \vert 1\rangle_1 \vert
0 \rangle_2  + r_2 \vert 0\rangle_1 \vert 1 \rangle_2.
\end{eqnarray}
The function ${\cal
E}(t)$ is the survival amplitude of the superradiant state
$\langle \psi_+ (t) \vert \psi_+ (0) \rangle = {\cal E}(t)$.
If we express the initial state of the qubits
as a superposition of $\vert \psi_{\pm} \rangle$, that is $\vert
\psi (0) \rangle = \beta_- \vert \psi_- \rangle + \beta_+ \vert
\psi_+ \rangle$ with $\beta_{\pm}= \braket{\psi_{\pm}}{\psi(0)}$,
we see that, while part of the initial state will be trapped in
the subradiant state $\vert \psi_- \rangle$, another part will
decay following Eq.~(\ref{eq:E}). Thus the amount of entanglement that survives
depends on the specific initial state and on the value of the coefficients $r_j$.

\subsection{Non-subradiant Scenario}
For $\omega_1 \ne \omega_2 $ no subradiant or decoherence-free
state exists and, as a consequence, the analytical expression for
the amplitudes $c_{1,2}(t)$ becomes more complicated
\begin{eqnarray}
c_1(t)&=&  {\cal E}_{11}(t;r_1) c_1(0) + {\cal E}_{12}(t;r_1) c_2(0), \label{eq:c1Sc2} \\
c_2(t)&=&  {\cal E}_{21}(t;r_1) c_1(0) + {\cal E}_{22}(t;r_1)
c_2(0), \label{eq:c2Sc2}
\end{eqnarray}
where the functions ${\cal E}_{ij}(t;r_1)$ depend not only on time but also on the value of $r_1$.

We emphasize that in both scenarios, the solution of the
differential equations for the amplitudes $c_{1,2}(t)$ is exact as
we have not performed neither the Born nor the Markov
approximation. The structure of the functions ${\cal
E}_{ij}(t;r_1)$ and the main steps to the solution are briefly
outlined in Appendix A.

\subsection{Dispersive regime}
In this subsection we focus on the system dynamics when the qubits
are far off-resonant from the main cavity mode, i.e. for
$\delta_1, \delta_2 \gg {\cal R}$. In this regime, both in the
subradiant and in the non-subradiant scenarios, the main features
of the dynamics can be obtained by looking at the effective
dispersive Hamiltonian describing the coupling of the two qubits
with a single-mode cavity field \cite{zheng00,blais04,rao} and
remembering that this behavior must then be corrected taking into
account the effect of the cavity losses. In Appendix B we derive
the effective dispersive Hamiltonian for this system, assuming
that the cavity field is initially in the vacuum state,
\begin{equation}
\label{Heff-SubSc} H_{eff}= \sum_{j=1}^2\frac{\mathcal{R}^2 \,
r_j^2}{\delta_j}\,
\sigma_+^{(j)}\sigma_-^{(j)}+\frac{\mathcal{R}^2 \, r_1 r_2}{2 \,
\delta_j}\left(\sigma_+^{(1)}\sigma_-^{(2)}+\sigma_+^{(2)}\sigma_-^{(1)}\right).
\end{equation}

The first two terms in the Hamiltonian are proportional to
$\sigma_+^{(j)} \sigma_-^{(j)}$ and describe the Stark shifts  due
to the dispersive interaction with the cavity vacuum. The
remaining terms describe an effective dipole-dipole coupling
between the two atoms induced by the cavity mode. As we will see in
the following these two terms play an essential role in the
entanglement generation process. By looking at Eq.
(\ref{Heff-SubSc}) we notice that both the Stark shifts and the
effective interaction strength between the qubits are now $\propto
\mathcal{R}^2/\delta_{1,2}$.

In the dispersive regime  the cavity is only virtually excited,
thus the photon loss is less important and the effective
decoherence rate due to the cavity decay is strongly suppressed to
the advantage of the generation of entanglement. As we will see in
Sec. III A for the subradiant scenario, the effective decoherence
rate due to the cavity decay in this case becomes $(\mathcal{R}^2
/\delta^2) \lambda$.

\section{Entanglement Dynamics}
To study the time evolution of the two-qubit entanglement we use
the concurrence $C(t)$ \cite{wootte}. This is an entanglement
measure related to the entanglement of formation, ranging from one
for maximally entangled states to zero for separable ones.

For the system of two qubits described by the reduced density matrix
of Eq.~(\ref{eq:rhos}) the concurrence takes a very simple form
\begin{equation}
\label{eq:cdef} C(t) = 2 \left| c_1(t)\, c_2^*(t) \right| .
\end{equation}
Such equation shows a relation between the behavior of the
concurrence and the time evolution of the excitation shared by the
two qubits. Having  in mind the considerations of Sec. I D one may
understand how, through a suitable choice of the detuning between
the qubits and the cavity, it is possible to improve both the
generation of entanglement and its preservation for long times.

To better discuss the time evolution of the concurrence as a
function of the initial amount of entanglement stored in the
system, we consider a general initial states of the form given by Eq.
(\ref{initialstate})
with
$$c_{01}= \sqrt{\frac{1-s}{2}}, \quad c_{02}= \sqrt{\frac{1+s}{2}} \,
e^{i \phi}, \ \mbox{with}\quad  -1 \le s \le 1. $$ Here, the separability
parameter $s$ is related to the initial concurrence as
$s^2=1-C(0)^2$.

Before describing in detail the dynamics in the subradiant
scenario (Sec. III) and non-subradiant scenario (Sec. IV), it is
useful to recall the main features of the time evolution of the
entanglement when $\omega_1=\omega_2=\omega_c$, i.e. in the
resonant case, as discussed in Ref. \cite{Man08}.

(i) The concurrence dynamics, as well as the value of the
stationary concurrence, depends on the relative coupling strength,
i.e. on the parameter $r_1$.

(ii) For certain entangled initial states, there exist at least
one time instant $\bar{t} < \infty$ at which $C(\bar{t}) = 0$,
both  in the strong coupling (good cavity) and weak coupling (bad
cavity) limits, i.e. for $\lambda \ll \mathcal{R}$ and $\lambda
\gg \mathcal{R}$, respectively [See Fig. \ref{fig:SdBadCavS0}
(a)].

(iii) In the weak coupling (bad cavity) limit, for an initially
factorized state, the reservoir creates entanglement and this is
indicated by a monotonic increase in the value of the concurrence.

(iv) In the weak coupling (bad cavity) limit, for an initially
entangled state, the reservoir causes entanglement loss and the
concurrence decreases with time until reaching, in some cases, the
value zero, after which a small fraction of entanglement can be
recreated [See Fig. \ref{fig:SdBadCavS0} (a)].

(v) In the strong coupling (good cavity) limit, oscillations in
the concurrence appear. For an initially factorized state there
exist times at which the value of the concurrence is higher than
the value of the stationary concurrence.

In the next two sections we are going to study how the time
evolution of the concurrence is modified in presence of detuning.

\section{Off-resonant Entanglement in the Subradiant Scenario}

We begin considering the case $\omega_1=\omega_2$. Whenever
possible, rather than discussing the exact expression of the
concurrence, we will try to derive simpler approximated
expressions which are useful for understanding the physical processes taking
place in the system.

\subsection{Bad cavity limit - Enhancement of the entanglement generation}

\begin{figure}[!htb]
\includegraphics[width=0.48\textwidth]{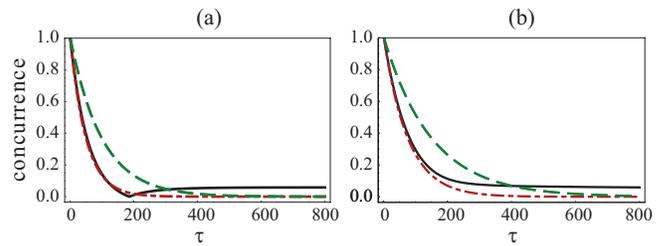}
\caption{(Color online) Time evolution of the  concurrence in the
bad cavity limit ($R=0.1$) with $s=0$ and $\phi=0$, for the cases
of i) maximal stationary value, when $r_1=\sqrt{3}/2$
(black solid line), ii) symmetrical coupling $r_1=1/\sqrt{2} $
(red dot-dashed line), and iii) only one coupled atom $r_1=0,1$
(green dashed line). For each of such cases, we describe the
entanglement dynamics in two different coupling regime: the
resonant limit (left plot) and for
$\delta_1=\delta_2 = 0.7 \lambda$ (right plot).}
\label{fig:SdBadCavS0}
\end{figure}

In the bad cavity case, e.g., for $R= \mathcal{R}/\lambda= 0.1$,
and for small values of the detuning $\delta < \mathcal{R}$, the
behavior of the concurrence does not change appreciably compared
to the resonant case. For values of the detuning $\delta \approx
\mathcal{R}$, i.e. when approaching the dispersive regime, the
dynamics for an initially factorized state ($s=1$) shows a
monotonic increase towards the stationary value of the concurrence
as in the resonant case. However, a significant change occurs in
the bad cavity limit when the system is prepared in an initial
entangled state. Indeed one can prove that in this regime,
contrary to the resonant case, a finite time $\bar{t}$ such that
$C(\bar{t}) = 0$ [See Fig. \ref{fig:SdBadCavS0} (b)] does not exist
anymore.

We now focus on the dispersive regime $\delta \gg \lambda \gg
\mathcal{R}$. If the qubit-system is initially entangled, e.g.,
for $s=0$, the expression for the concurrence can be simplified as
follows
\begin{eqnarray}
  C(t)=|{\cal E}| &\approx & e^{-\frac{\mathcal{R}^2}{\delta^2} \lambda t}, \qquad \mathrm{for} \quad r_1=0,1;\\
  C(t)=|{\cal E}|^2 &\approx & e^{-2 \frac{\mathcal{R}^2}{\delta^2} \lambda t}, \qquad
  \mathrm{for}\quad  r_1=1/\sqrt{2}.
\end{eqnarray}
The equations above show that the concurrence vanishes with the decay
rate $(\mathcal{R}^2/\delta^2) \lambda$ when only one of the two
qubits is coupled to the environment ($r_1=0,1$), and with 
$2(\mathcal{R}^2/\delta^2) \lambda$  when both qubits are
identically coupled to the environment (for $r_1=1/\sqrt{2}$).
Since $\mathcal{R}/\delta \ll 1$ this proves that in the
dispersive regime the decay of entanglement is strongly inhibited
compared to the resonant regime since in this case the two atoms
exchange energy only via the virtual excitation of the cavity
field and therefore the cavity losses do not affect strongly the
dynamics.

\begin{figure}[!htb]
\includegraphics[width=0.48\textwidth]{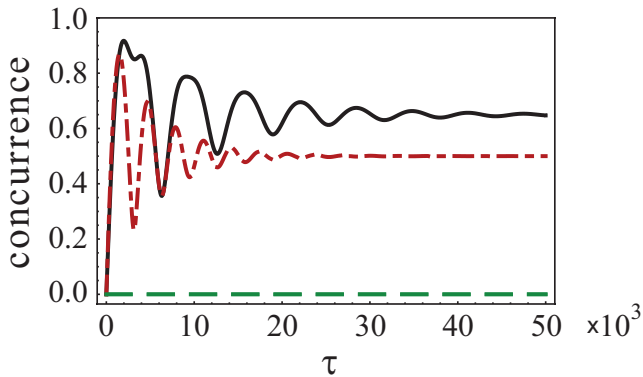}
\caption{(Color online) Time evolution of the  concurrence in the
bad cavity limit ($R=0.1$) with $s=1$, for the cases of i) maximal
stationary value, $r_1=\sqrt{3}/2$ (black solid line), ii)
symmetrical coupling $r_1=1/\sqrt{2} $ (red dot-dashed line), and
iii) only one coupled atom $r_1=0,1$ (green dashed line). All of
the plots describe the dispersive regime with $\delta_1=\delta_2 =
10 \lambda$.} \label{fig:SdBadCavS1}
\end{figure}

For large enough detunings the entanglement shows oscillations as
a function of time for all of the initial atomic states for which
a finite stationary concurrence is obtained, $C_s \neq 0$. Due to
the presence of these oscillations and for an initially factorized
state,  the concurrence reaches values greater than the stationary
value $C_s$ even in the bad cavity limit, as shown in Fig.
\ref{fig:SdBadCavS1}. For example, for $r_1=\sqrt{3}/2$, $R=0.1$
and $\delta=10 \lambda$, at $\lambda t \approx 2 \times 10^3$ the
concurrence reaches the value $C=0.92$. For an initially
factorized state ($s=1$) and for $r_1=1/\sqrt{2}$ we can derive
the following approximated expression for the concurrence
\begin{equation}\label{eq:ConDispS1}
    C(t) \approx \frac{1}{2}\sqrt{1+e^{-4 \frac{\mathcal{R}^2}{\delta^2}\lambda t}
    -2 e^{-2\frac{\mathcal{R}^2}{\delta^2}\lambda t}\cos\left( 2 \frac{\mathcal{R}^2}
    {\delta}t\right)}.
\end{equation}
From this equation one sees that $C(t)$ attains its maximum value at $t = \frac{\pi \delta}{2
\mathcal{R}^2}$. This formula also shows that the concurrence undergoes a
series of damped oscillations with frequency $2
\mathcal{R}^2/\delta$ and decay rate $2(\mathcal{R}/\delta)^2
\lambda$.


With increasing detuning, the oscillations become more and more
regular, quasi-periodic. The pattern is similar to the oscillations characterizing
the strong coupling regime, but now the period is longer.
As we will see in Sec. III B, the generation of a high degree of
entanglement in the dispersive regime for initially separable
state can be achieved also in the good cavity limit. However it is
remarkable that already in the bad cavity limit, values of
concurrence close to one can be generated. Our approach
generalizes the results obtained for the dispersive regime in Ref.
\cite{zheng00} in the ideal cavity limit to the more realistic
case of cavity losses.

\subsection{Good cavity limit - Entanglement quantum beats}

\begin{figure}[!htb]
\includegraphics[width=0.48\textwidth]{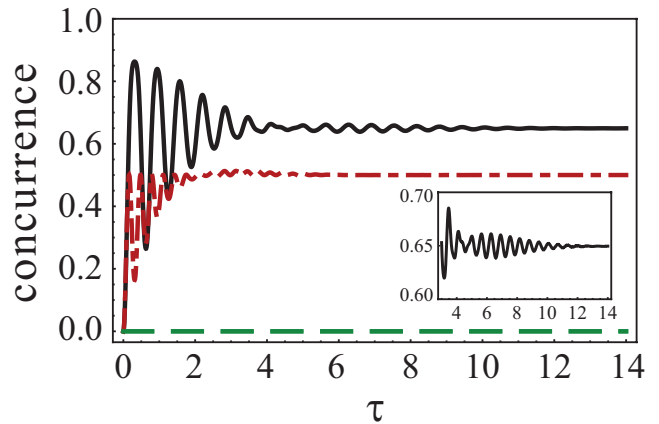}
\caption{(Color online) Time evolution of the  concurrence in the
good cavity limit ($R=10$) with $s=1$, for the cases of i) maximal
stationary value $r_1=\sqrt{3}/2$ (black solid line), ii)
symmetrical coupling $r_1=1/\sqrt{2} $ (red dot-dashed line), and
iii) only one coupled atom $r_1=0,1$ (green dashed line). The
curves are drawn for small detuning, $\delta_1=\delta_2 = 0.7
\lambda$; thus, outside the dispersive region. The inset shows the
entanglement beat for the case i).} \label{fig:SdGoodCavS1}
\end{figure}

In the strong coupling case entanglement oscillations are present
for any initial atomic state. Moreover, for $\delta \approx
\lambda \ll \mathcal{R}$, when both atoms are effectively coupled
to the cavity field, i.e., $r_1 \neq 0,1$, the dynamics of the
concurrence is characterized by the occurrence of quantum beats,
as shown in Fig. \ref{fig:SdGoodCavS1}. For initially entangled
states this phenomenon is more evident for $\phi=\pi$ because the
value of stationary entanglement in this case is higher and the
behavior of the concurrence is more regular.

In order to better understand the origin of these entanglement
beats, we consider the case
 $s=1$ and $r_1=1/\sqrt{2}$. For these values of the parameters, and for $\delta \approx \lambda \ll
\mathcal{R}$, the expression of the concurrence can be written as follows,
\begin{equation}\label{eq:ConSmDetS1}
    C(t) \approx \frac{1}{2}\sqrt{1+e^{-2 \lambda t} \cos(\mathcal{R} t)^4
    -2 e^{- \lambda t} \cos(\mathcal{R} t)^2 \cos(\delta t)}.
\end{equation}
The
term
$$
\cos(\mathcal{R} t)^2 \cos(\delta t)=\frac{1}{2}\cos(\delta t) \left[1+\cos(2\mathcal{R}
t)\right]
$$
in Eq. (\ref{eq:ConSmDetS1}),  describing an oscillation at
frequency $2 \mathcal{R}$   modulated by a slower one with frequency
 $\delta$, is responsible for the occurrence of the
quantum beats.

To gain insight in the physical processes characterizing the
dynamics, we consider  the energy spectrum of  the dressed states
in the off-resonant case but in the absence of damping, as shown
in Fig.~\ref{fig:DressLev2Q}. The diagonalization of the
Tavis-Cummings Hamiltonian [See Eq. (\ref{HamFeno}) in Appendix B]
yields the dressed states
\begin{eqnarray}
\ket{\phi_+} &=&
\frac{1}{\sqrt{\omega_-^2+\mathcal{R}^2}}\left(-\mathcal{R}\ket{\psi_+}\ket{0}_R+
\omega_-\ket{00}\ket{1}_R \right), \nonumber \\     \\
\ket{\phi_-} &=&
\frac{1}{\sqrt{\omega_+^2+\mathcal{R}^2}}\left(-\mathcal{R}\ket{\psi_+}\ket{0}_R+
\omega_+\ket{00}\ket{1}_R \right), \nonumber \\      \\
\ket{\phi_0} &=&
\ket{\psi_-}\ket{0}_R.
\end{eqnarray}
The corresponding eigenenergies  are given by
\begin{eqnarray}
  \omega_\pm &=& \frac{1}{2}\left( \delta \pm \sqrt{4 \mathcal{R}^2+\delta^2}\ \right), \\
  \omega_0 &=& \delta ,
\end{eqnarray}
where $\mathcal{R}= g \alpha_T$ is the vacuum Rabi frequency and
$\delta$ is the qubits-cavity detuning.

\begin{figure}[!htb]
\includegraphics[width=0.45\textwidth]{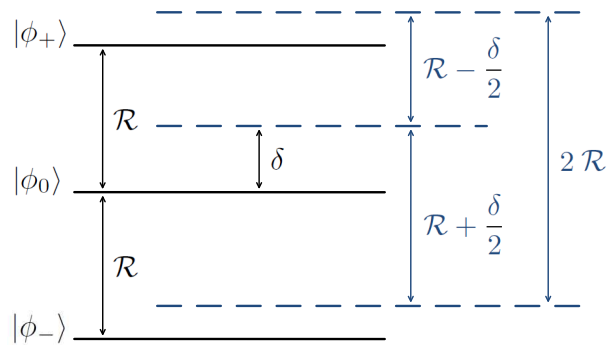}
\caption{(Color online) Energy spectrum of the dressed
qubit-photon states in the case of small detuning (blue dashed
line) and in the resonant coupling case (black solid line).}
\label{fig:DressLev2Q}
\end{figure}
\begin{figure}[!htb]
\includegraphics[width=0.48\textwidth]{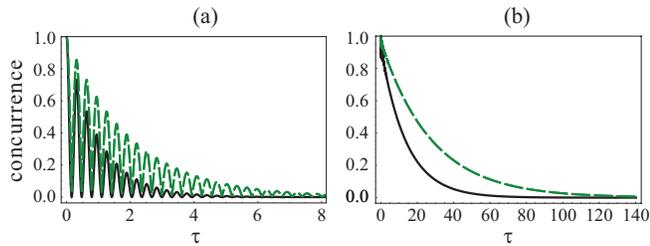}
\caption{(Color online) Time evolution of the  concurrence in the
good cavity limit ($R=10$), with $s=0$ and $\phi=0$, for the cases
of i) symmetrical coupling $r_1=1/\sqrt{2} $ (black solid line),
and ii) only one coupled atom $r_1=0,1$ (green dashed line). The
two plots describe two different detuning regions:
$\delta_1=\delta_2 =0.7 \lambda$ (left) and $\delta_1=\delta_2 =
50 \lambda$ (right).} \label{fig:SdGoodCavS0}
\end{figure}

On the other hand, the unperturbed states can be expressed as a
superposition of the $\ket{\phi_\pm}$,$\ket{\phi_0}$ eigenstates,
with probability amplitudes evolving at frequencies $\omega_\pm$
and $\omega_0$. The effect of the detuning is a shift of the
qubits-cavity energy levels, thus the qubits-field coupling gives
rise to a reversible energy exchange between unperturbed state at
frequencies $2 \mathcal{R}$, $\mathcal{R}-\delta/2$ and
$\mathcal{R}+\delta/2$. This is clearly seen, e.g., from the time
evolution of the populations
\begin{eqnarray*}
  &&|c_2(t)|^2= |\bra{01}\bra{0}e^{-i H t}\ket{01}\ket{0}|^2 \\
   && \qquad =  r_1^4+\frac{r_2^4}{2}\left[ 1+\cos(2 \mathcal{R} t) \right]+2 r_1^2
   r_2^2 \cos \left(\mathcal{R} t \right)\cos \left(\frac{\delta}{2}t \right).
\end{eqnarray*}
The equation above contains a term oscillating at frequency $2
\mathcal{R}$, coming from the coupling between the dressed states
$\ket{\phi_+}$ and $\ket{\phi_-}$,  and a term oscillating at
frequency $\mathcal{R}$ modulated by $\delta$ coming from the
interference between the oscillations  at frequencies
$\mathcal{R}-\delta/2$ and $\mathcal{R}+\delta/2$ that couple the
states $\ket{\phi_+}$-$\ket{\phi_0}$ and
$\ket{\phi_-}$-$\ket{\phi_0}$, respectively.

In the discussion above we have disregarded the cavity losses.
When they are taken into account one sees that the dressed energy
splitting is resolved, and therefore the quantum beats will be
visible, if $2\mathcal{R}$ is larger than the decay width
$\lambda$. This is achieved in strong coupling regime. Therefore,
one does not observe quantum beats in the bad cavity case.

We conclude this section studying how the detuning influences the
decay of entanglement, for an initially maximally entangled state
of the system, and the reservoir-induced entanglement generation,
for an initial factorized state.

When only one of the two qubits is effectively coupled to the
cavity field, i.e. for $r_1=0,1$, for maximally entangled initial
states ($s=0$) in the resonant regime, $\delta =0$, the system
performs damped oscillations between the states $\vert \psi_+
\rangle$ and $\vert \psi_- \rangle$, which are equally populated
at the beginning. Hence entanglement revivals with maximum
amplitude are present in the dynamics, as shown in Fig.
\ref{fig:SdGoodCavS0} (a). Increasing the detuning, the amplitude
of the oscillations decreases and the revivals disappear, while
the frequency does not change appreciably, [See Fig.
\ref{fig:SdGoodCavS0} (a)].
In this case the expression of the concurrence for small values of
the detuning can  be written as
\begin{widetext}
\begin{equation}\label{eq:ConSmaDetS0r0}
C(t)=|{\cal E}| \approx e^{-\lambda t / 2} \sqrt{\cos(\mathcal{R}
t)^2+\frac{\delta^2 + \lambda^2}{4 \mathcal{R}^2}\sin(\mathcal{R}
t)^2-\frac{\lambda}{\mathcal{R}}\sin(\mathcal{R}
t)\cos(\mathcal{R} t)},
\end{equation}
\end{widetext}
while for greater values of the detuning, the oscillations
completely disappear and the concurrence decays exponentially
\begin{equation}\label{eq:ConFarDetS0r0}
    C(t)=|{\cal E}| \approx e^{-\frac{\mathcal{R}^2}{\delta^2}\lambda t},
\end{equation}
as shown in Fig. \ref{fig:SdGoodCavS0} (b).

Finally, we note that, similarly to the behavior discussed in the
bad cavity limit, when the qubits are initially in a factorized
state, the presence of the detuning enhances the generation of
entanglement at short times compared to the resonant coupling
case, as illustrated in Fig. \ref{fig:SdBadCavS1}. In general, in
the strongly dispersive regime, the qubits do not exchange energy
with the cavity, which is only virtually excited. Thus a high degree
of reservoir-induced entanglement can be generated  both in the
good and in the bad cavity limits.

\section{Off-resonant Entanglement in the non-subradiant scenario}
In this section, we analyze the more general situation in which
the transition frequencies of the qubits are different, $\omega_1
\neq \omega_2$, and both qubits are off-resonant with the cavity
field. Due to the absence of a subradiant state, even a small
value of the detunings $\delta_1,\delta_2 \ll \mathcal{R}$
contributes to accelerate the decay of entanglement for every
initial states. For an initially factorized state, in the bad
cavity limit, the entanglement initially created via the
interaction with the reservoir is rapidly destroyed as time
evolves. In the good cavity limit entanglement oscillations are
present and also quantum beats of entanglement can be observed for
$\delta_1, \delta_2 \approx \lambda \ll \mathcal{R}$.

\begin{figure}[!htb]
\includegraphics[width=0.48\textwidth]{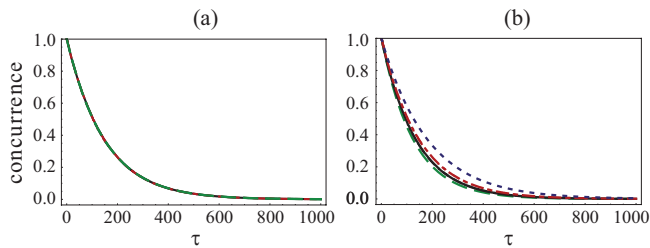}
\caption{(Color online) Time evolution of the  concurrence in the
bad cavity limit ($R=0.1$), with $s=0$ and $\phi=0$, for the cases
of i) maximal stationary value, corresponding to 
 $r_1=\sqrt{3}/2$ (black solid line), ii) symmetrical
coupling $r_1=1/\sqrt{2} $ (red dot-dashed line), and iii) only
one coupled atom $r_1=1$ (green dashed line) and $r_1=0$ (blue
dotted line). Two different detuning pairs are represented: the
symmetrical detuning with $\delta_1 =-0.7 \lambda$, $\delta_2 =0.7
\lambda$ (left plot) and the asymmetrical detuning with $\delta_1
=-0.5 \lambda$, $\delta_2 =0.9 \lambda$ (right plot).}
\label{fig:SyAsyBad}
\end{figure}

We now consider in more detail the case in which the two qubits
frequencies are symmetrically detuned from the central peak of the
Lorentzian spectrum describing the field inside the cavity. In the
dispersive region $\delta \gg \mathcal{R}$, and for initially
entangled states ($s=0$), the concurrence vanishes without
manifesting a dominant dependence from $r_1$ and $\phi$. In other
words, all the states initially entangled decay following the same
behavior in such regime, as shown in Fig. \ref{fig:SyAsyBad} (a).
This is in contrast to what we observed in all other regimes,
where a dependence on the value of $r_1$ is present. We stress
once more that this feature seems to occur only for the case of
symmetric detuning. Indeed, when introducing a small asymmetry in
the value of the detunings the behavior of the concurrence shows
again a dependence on the parameter $r_1$, as illustrated in Fig.
\ref{fig:SyAsyBad} (b).

In order  to understand the peculiar behavior of the concurrence
in the dispersive regime and for symmetric detunings we once more
start by neglecting the effect of the cavity losses and use the
dispersive Hamiltonian given by Eq. (\ref{Heff-SubSc}). For
symmetrical detunings $\delta_1 = - \delta_2$ this equation takes
the form
\begin{equation}\label{eq:HeffSimm}
H_{eff}=-\frac{\mathcal{R}^2
r_1^2}{\delta}\sigma_+^{(1)}\sigma_-^{(1)}+\frac{\mathcal{R}^2
r_2^2}{\delta}\sigma_+^{(2)}\sigma_-^{(2)},
\end{equation}
with $\delta=|\delta_1|=|\delta_2|$.

Comparing Eq. (\ref{eq:HeffSimm}) with Eq. (\ref{Heff-SubSc}) we
notice that the terms describing the effective dipole-dipole
coupling induced by the cavity mode are here absent. Therefore the
only remaining effect is the entanglement decay induced by the
cavity losses. The decay rate, however, does not depend on the
relative coupling parameter $r_1$ but only on the total coupling
strength $\alpha_T$ via the vacuum Rabi frequency $\mathcal{R}$.
This explains why, even when the cavity losses are taken into
account, the time evolution of the concurrence for symmetric
detunings does not depend on $r_1$. When a small asymmetry in the
detunings is introduced, the dipole-dipole effective coupling
terms are non-zero and, due to the presence of $r_1$ and $r_2$ in
the effective dipole-dipole coupling strength, the dynamics
becomes again dependent on $r_1$.

\section{Summary and Conclusions}
In this paper we have provided a complete analysis of the exact
dynamics of the entanglement for two qubits interacting with a
common zero-temperature reservoir in the off-resonant case. We
have presented a general analytical solution for the two-qubit
dynamics without performing the Born-Markov approximation. In the
case of a Lorentzian spectrum, describing, e.g., the
electromagnetic field inside a single mode lossy cavity, we have
obtained explicit expressions for the reduced density matrix and
for the concurrence. The availability of the exact solution
allowed us to look at the entanglement dynamics both in the weak
coupling (bad cavity)  and in the strong coupling (good cavity)
limits.

If the two qubits are initially disentangled, the interaction with
the common reservoir generates entanglement. Our results
demonstrate that a high degree of entanglement can be generated in
this way, especially in the dispersive regime, and even in the bad
cavity limit. For initially entangled states, the concurrence
decay is slowed down when the qubits are detuned from the peak of
the Lorentzian. In this case, indeed, the cavity losses affect
less the atoms dynamics since the effective atom-atom interaction
is mediated by virtual photon exchange.

In general, the entanglement dynamics is strongly sensitive to the
relative coupling parameter $r_1$, indicating how strongly each of
the two qubit is individually coupled to the e.m. field. Only when
the qubits frequencies are symmetrically detuned from the main
cavity frequency, in the dispersive regime, the dependence on the
relative coupling disappears. Finally we have discovered that, in
the strong coupling regime, for intermediate values of the
detuning, the dynamics of the concurrence shows the occurrence of
quantum beats. We have given a physical interpretation of this
phenomenon in terms of the quantum interference between the
transitions among the dressed states of the atomic system.

We believe that our results contribute in shedding light on the
behavior of quantum entanglement in realistic conditions, that is
when the effect of the environment on the quantum system is taken
into account. For this reason they have both a fundamental and an
applicative value and they indicate how rich the dynamics of this
system can be. The model we have studied can be employed to
describe both trapped ions in optical cavities \cite{Walther} and
circuit cavity QED dynamics \cite{wallra,mika,wallra2}. In both
physical contexts, the observation of the effects we have
discussed should be achievable with the current experimental
technologies.

\section{Acknowledgements}
S.M., K.-A.S., and J.P. acknowledge financial support from the Academy of Finland
Projects No.~108699, No.~115682 and No.~115982, the Magnus Ehrnrooth
Foundation, and the V\"ais\"al\"a Foundation.  F.F. acknowledges financial support from the CIMO Project No. TM-08-5425 and
thanks N. Lo Gullo and R. L. Zaffino for useful discussions, and
S.M. and all of the Quantum Optics Group members at
the University of Turku for the kind hospitality. 
Finally, the authors acknowledge W. Lange and B. Garraway for discussions on the experimental implementation with ion-cavity QED set up.

\section*{Appendix A - Analytic solution for the probability amplitudes}
In this Appendix we briefly discuss the structure of the
analytical solutions of Eqs.~(\ref{eq:Dotc1f})-(\ref{eq:Dotc2f})
for the probability amplitudes  $c_{1,2}(t)$ and how they can be
obtained applying the Laplace transform method. We note that the
solutions obtained in this way are exact since we do not perform
any kind of approximation.

The solution of the Laplace transformed amplitudes
$\widetilde{c}_{1,2}(s)$, obtained from
Eqs.~(\ref{eq:LapTrc1f})-(\ref{eq:LapTrc2f}) can be written as the
sum of three ratios having denominators $(s-s_i)$, where $s_i$ are
the roots of the cubic equation.
\begin{equation}\label{CubEq}
    s^3+ A_j s^2 + B_j s + C_j = 0, \qquad (j=1,2)
\end{equation}
where
\begin{eqnarray*}
  A_{1,2} &=& \lambda +i\, \left(\delta_{2,1} - 2\, \delta_{1,2}\right), \\
  B_{1,2} &=& \mathcal{R}^2 - \delta_{1,2}^2 + \delta_1 \, \delta_2 +i \left(\delta_{2,1}
    -\delta_{1,2}\right)\, \lambda, \\
  C_{1,2} &=& i \, \mathcal{R}^2 \, r_{1,2}^2 \, \left(\delta_{2,1} -\delta_{1,2}\right).
\end{eqnarray*}
The amplitudes $c_{1,2}(t)$, obtained by inverse Laplace transform will then be the sum of
three damped oscillating terms having, in general, a
complicated structure.
Only for the case  $\omega_1=\omega_2$ a simple
analytical expressions for the probability amplitudes can
be obtained, whereas in the general case there is no simple
solution. This is because, when  $\omega_1=\omega_2$ the cubic equation can be
written as a product of polynomials of first and second order having always one
root coincident with zero. In this case one can write the
amplitudes in the simple form given by Eqs.~(\ref{eq:c1Sc1})-(\ref{eq:c2Sc1}).

\section*{Appendix B - Effective dispersive Hamiltonian }

The Hamiltonian describing the interaction between two-qubit
systems and the quantized cavity mode is given by
\begin{eqnarray}\label{HamFeno}
\nonumber  H &=& \sum_{j=1}^2\omega_j \sigma^{(j)}_+  \sigma^{(j)}_- + \omega_c b^{\dag} b \\
\nonumber   & & + \left[g \left( \alpha_1 \sigma^{(1)}_+ + \alpha_2\sigma^{(2)}_+ \right)  b  + {\rm h.c.} \right].
\end{eqnarray}

To obtain the effective Hamiltonian describing the interaction
with the cavity in the dispersive regime, one can apply  the
canonical transformation defined by the unitary
operator \cite{blais04,rao}
\begin{equation}\label{eq:TranCan}
   e^{\alpha S}= e^{-\sum_{j=1}^2\frac{\mathcal{R} r_j}{\delta_j}\left(b \, \sigma_+^{(j)} -b^\dag
\sigma_-^{(j)}\right)}
\end{equation}
with $\mathcal{R} r_j= g \alpha_j$. This procedure is correct to the
second order in the coupling to the cavity, and, limiting
ourselves to this approximation, we can write the effective
Hamiltonian as follows
$$ H_{eff} = e^{\alpha S}H e^{-\alpha
S}\simeq H+\alpha[S,H]+\frac{\alpha^2}{2}[S,[S,H]].
$$
Assuming that the cavity field is initially in the vacuum state,
$H_{eff}$ takes the form
\begin{equation}\label{EffHam}
    H_{eff}= \sum_{j=1}^2\frac{\mathcal{R}^2 \, r_j^2}{\delta_j}\, \sigma_+^{(j)}\sigma_-^{(j)}+\frac{\mathcal{R}^2 \, r_1
    r_2}{2 \, \delta_j}\left(\sigma_+^{(1)}\sigma_-^{(2)}+\sigma_+^{(2)}\sigma_-^{(1)}\right),
\end{equation}
where the terms proportional to $\sigma_+^{(j)} \sigma_-^{(j)}$
describe the Stark shifts due to the dispersive interaction, while
the last two terms describe the dipole-dipole coupling between the
two atoms induced by the cavity mode through the exchange of
virtual cavity photons.

\section*{Appendix C - Approximate expressions of the concurrence}
In this Appendix we derive approximate expressions for the amplitudes
$c_{1,2}(t)$ in the case of large (and equal) detuning $\delta \gg
\lambda \gg \mathcal{R}$.

For this purpose, we expand the term $\Omega= \sqrt{\lambda^2 -
\Omega_R^2-i 2 \delta \lambda}\,$ as follows,
\begin{equation}\label{OmefarDet}
    \Omega \approx \lambda \left(1- \frac{2 \, \mathcal{R}^2}{\delta^2}
    \right)-i \left( \delta +\frac{2 \, \mathcal{R}^2}{\delta} \right).
\end{equation}
The temporal evolution described by $\mathcal{E}(t)$ can then be
written as
\begin{eqnarray*}
\mathcal{E}(t) & \approx & e^{-(\lambda -i
\delta)t/2}\left[\cosh\left(\frac{\Omega
t}{2}\right)+\sinh\left(\frac{\Omega t}{2}\right)\right] \\
& \approx & e^{-\frac{\mathcal{R}^2}{\delta^2}(\lambda +i
\delta)t}.
\end{eqnarray*}

For the sake of simplicity we consider here the case $s=1$ and
$r_1=1/\sqrt{2}$. However  the time evolution of the concurrence
has features in  common with all of the other cases:
\begin{eqnarray*}
  C(t) &=& 2 \, \vert c_1(t) \vert \, \vert c_2(t)\vert \\
   &=& \frac{1}{2}\sqrt{\left(1+\vert \mathcal{E}(t)\vert^2\right)^2-\left(2 \, Re[\mathcal{E}(t)]\right)^2} \\
   &\approx & \frac{1}{2}\sqrt{1+e^{-4 \frac{\mathcal{R}^2}{\delta^2}\lambda t}
    -2 e^{-2\frac{\mathcal{R}^2}{\delta^2}\lambda t}\cos\left( 2 \frac{\mathcal{R}^2}
    {\delta}t\right)}.
\end{eqnarray*}

On other hand, for small detunings of the order of $\lambda$,
outside the dispersive region $\delta \ll \mathcal{R}$, the
approximate form of $\Omega$ is given by
\begin{equation}\label{OmefarDet}
    \Omega \approx  \frac{\lambda \, \delta}{2 \, \mathcal{R}} -i 2 \, \mathcal{R},
\end{equation}
so that the time evolution is described by the function
\begin{eqnarray*}
\mathcal{E}(t) & \approx & e^{-(\lambda -i
\delta)t/2}\left[\cosh\left(\frac{\Omega t}{2}\right)-\frac{\delta
+ i \lambda}{2 \, \mathcal{R}}\sinh\left(\frac{\Omega
t}{2}\right)\right] \\
& \approx & e^{-(\lambda -i
\delta)t/2}\left[\cos\left(\mathcal{R} t \right)-\frac{\lambda}{2
\, \mathcal{R}}\sin\left(\mathcal{R} t \right)+i \frac{\delta}{2
\, \mathcal{R}}\sin\left(\mathcal{R} t \right) \right].
\end{eqnarray*}
Therefore, for the case $s=1$ and $r_1=1/\sqrt{2}$ the time
evolution of the concurrence is given by
\begin{eqnarray*}
  C(t) &=& \frac{1}{2}\sqrt{(1+\vert \mathcal{E}(t)\vert^2)^2-(2 \, Re[\mathcal{E}(t)])^2} \\
   &\approx & \frac{1}{2}\sqrt{1+e^{-2 \lambda t} \cos(\mathcal{R} t)^4
    -2 e^{- \lambda t} \cos(\mathcal{R} t)^2 \cos(\delta t)}.
\end{eqnarray*}

\end{document}